\documentclass[pra,a4paper,twocolumn,showpacs,superscriptaddress]{revtex4}
\usepackage{amsmath}
\usepackage{amsfonts}
\usepackage{graphicx}
\usepackage{longtable}
\newcommand{\be}{\begin{equation}}
\newcommand{\ee}{\end{equation}}

\newcommand{\ba}[1]{\left(\begin{array}{#1}}
\newcommand{\ea}{\end{array}\right)}
\begin{document}

\title{Determining the whole pure symmetric $N$-qubit state from its parts} 

\author{A. R. Usha Devi}
\email{arutth@rediffmail.com}
\affiliation{Department of Physics, Bangalore University, 
Bangalore-560 056, India}
\affiliation{Inspire Institute Inc., Alexandria, 22303, USA.}
\author{Sudha}
\affiliation{Department of Physics, Kuvempu University, 
Shankaraghatta, Shimoga-577 451, India}
\author{A. K. Rajagopal} 
\affiliation{Inspire Institute Inc., Alexandria, 22303, USA.}

\date{\today}

\begin{abstract} 
The Majorana representation of 
symmetric $N$-qubit states is employed here to investigate how correlation 
information of the whole pure symmetric state gets imprinted in its parts. 
It is shown that reduced states of $(N - 1)$ qubits uniquely specify the entire class of pure 
$N$ qubit states containing two distinct spinors. 

\end{abstract}
\pacs{03.67.Mn, 03.67.-a}
\maketitle

\section{Introduction}

Knowing if higher order correlations in a multipartite system follow entirely from lower order ones involving fewer parties is one of the basic issues of interest in quantum information science~\cite{Niel}, as also in 
many body physics~\cite{Coleman}. Construction of the many electron state with the knowledge of its two particle reduced density matrices has been discussed in Ref.~\cite{M}. On the other hand, it is also well-known that $N$-party entanglement cannot, in general, be reversibly transformed into entanglement of two parties~\cite{Linden}. In this context,   Linden et. al.~\cite{SP}  proved an illuminating result that reduced states of a smaller fraction of the parties {\em uniquely} specify most of the {\em generic} multiparty pure states. Walck and Lyons~\cite{WL} showed that the $N$ party GHZ (Greenberger-Horne-Zeilinger) states and their local unitary equivalents are the only exceptions to this result. More recently, Preeti Parashar and Swapan Rana have shown that $N$ qubit W class states can be uniquely determined by their bipartite marginals~\cite{PS1}. They have also extended their arguments to determine generalized Dicke class states from their marginal density matrices~\cite{PS2}. 
 
In the present paper, we employ the Majorana representation~\cite{Majorana} to investigate the issue of determining the whole pure symmetric state in terms of its parts. We show that only {\em two} of the $N-1$ qubit reduced states determine the pure symmetric states characterized by two distinct Majorana spinors. We further show that our method offers a natural way of   determining a more general class of non-symmetric states too from their $N-1$ qubit reduced density matrices.  

The paper is organized as follows:  Majorana representation~\cite{Majorana} of $N$-qubit pure states obeying 
permutation symmetry and a classification~\cite{Solano} of symmetric states based on the distinct Majorana spinors constituting the state are discussed in Sec.~II.  In Sec.~III we focus on pure symmetric states of $N$ qubits,  containing two distinct spinors and show explicitly that the state  is uniquely determined by specifying only two of their $N-1$ party reduced states. We extend this method, in Sec.~IV, to prove that generalized states of the family of two distinct Majorana spinors are the unique {\em whole} pure states consistent with  their $N-1$ qubit subsystems. Sec.~V contains a summary of our results.

\section{Majorana representation for $N$-qubit pure symmetric states} 

Majorana~\cite{Majorana} expressed an arbitrary pure symmetric state $\vert \Psi_{\rm sym}\rangle$ of 
 spin $N/2$ systems as a  {\em symmetrized} combination  of  $N$ spinors as follows: 
\begin{equation}
\label{Maj}
\vert \Psi_{\rm sym}\rangle={\cal N}\, \sum_{P}\, \hat{P}\, \{\vert \epsilon_1, \epsilon_2, 
\ldots  \epsilon_N \rangle\} 
\end{equation} 
where 
\begin{eqnarray} 
\vert \epsilon_l\rangle&=&\cos(\beta_l/2)\, e^{-i\alpha_l/2}\, \vert 0\rangle
+\sin(\beta_l/2)\, e^{i\alpha_l/2}\, \vert 1\rangle,\nonumber \\
&&\hskip 0.8in   l=1,2,\ldots, N 
\end{eqnarray}
denote the spinors constituting the symmetric state $\vert \Psi_{\rm sym}\rangle$; $\hat{P}$ denotes the set of all $N!$ 
permutations and ${\cal N}$ corresponds to an overall normalization factor. Thus,  $N$ complex parameters 
\begin{equation}
z_l=\tan\frac{\beta_l}{2}e^{i\alpha_l}
\end{equation}  
 parametrize  the pure $N$ qubit symmetric state -- allowing for a  geometric representation of $N$ qubit pure symmetric states  in terms of  $N$ points on the Bloch sphere.    

On the other hand, states of $N$-qubits obeying exchange symmetry get restricted to 
a $(N+1)$ dimensional Hilbert space spanned by the collective basis vectors $\left\{\left\vert \frac{N}{2}, k-\frac{N}{2}\right\rangle, k=0,1,2,
\ldots N \right\}$ where,    
\begin{eqnarray}
\label{Dicke} 
\left\vert \frac{N}{2}, k-\frac{N}{2}\right\rangle &=&\frac{1}{\sqrt{^N C_k}}\,[\vert \underbrace{0, 0, \ldots}_{k}, 
\underbrace{1, 1, \ldots}_{N-k}\rangle \nonumber \\ 
&&\ \ \ +\ {\rm Permutations}\ ] 
\end{eqnarray}
are the $N+1$ Dicke  states -- expressed in the standard qubit basis $\vert 0\rangle,\ \vert 1\rangle$. (Here, 
 $^N C_k=\frac{N!}{k!\,(N-k)!}$ denotes the binomial coefficient). 
In other words, an arbitrary pure symmetric state,    
\begin{equation}
\label{sympure1} 
\vert \Psi_{\rm sym}\rangle =\sum_{k=0}^{N}\, c_k\, \left\vert \frac{N}{2}, k-\frac{N}{2}\right\rangle,
\end{equation}
is specified by the $(N+1)$ complex coefficients $c_k.$  Eliminating an overall phase and normalizing  the 
state  implies that $N$ complex parameters are required to 
completely characterize a pure symmetric state of $N$ qubits.

In order to express the coefficients $c_k$ in terms of the Majorana spinor orientations $(\alpha_l,\beta_l)$, 
it may be first identified that an identical rotation $R\otimes R\ldots \otimes R$ on the symmetric state $\vert \Psi_{\rm %%@
sym}\rangle$ transforms it into another symmetric state. Choosing $R=R_l^{-1}\equiv R^{-1}(\alpha_l,\beta_l,0)$ %%@
(where $(\alpha_l,\beta_l,0)$ denote the Euler angles of rotation~\cite{Rose})  aligning one of the %%@
constituent spinors say, $\vert \epsilon_l\rangle$, along the positive $z$-direction i.e., %%@
$R_l^{-1}\vert\epsilon_l\rangle=\vert 0\rangle$, results in the following identification, 
\begin{equation}
\label{rl}
\langle 1_1,1_2\ldots , 1_N\vert R_l^{-1}\otimes R_l^{-1}\ldots \otimes R_l^{-1}\, \vert \Psi_{\rm %%@
sym}\rangle\equiv 0
\end{equation}
This is because the rotation $R_l^{-1}\otimes R_l^{-1}\ldots \otimes R_l^{-1}$ takes one of the spinors $\vert \epsilon_l\rangle$ with orientation angles $(\alpha_l, \beta_l)$ to $\vert 0\rangle$ i.e., it aligns the spinor $\vert \epsilon_l\rangle$ in the positive $z$-direction. 
Then, every   term in the  superposition (\ref{Maj}) of the rotated state has atleast one 
$\vert 0\rangle$ and so, the projection $~\langle1_1,1_2,~\ldots,~ 1_N~\vert~ 
R_l^{-1}\otimes R_l^{-1}\otimes 
\ldots R_l^{-1}\vert \Psi^{(N)}_{\rm Sym}\rangle$ of the rotated state in the `all-down' direction
 vanishes. Eq.~(\ref{rl}) holds good for any identical rotations $R^{-1}_s\otimes 
R^{-1}_s\otimes\ldots\otimes 
R^{-1}_s,\ \ s=1,2,\ldots, N,$ orienting {\em any} one of the constituent  
qubits in the positive $z$-direction. In other words, there exist $N$ rotations 
$R^{-1}_s=R^{-1}(\alpha_s,\beta_s,0), s=1,2,\ldots , N$, which lead to the same result as in (\ref{rl}). 
 
In terms of the alternate representation (\ref{sympure1}), we obtain 
\begin{widetext}
\begin{eqnarray}
\label{cmj0}
 \langle N/2,-N/2\vert 
R_l^{-1}\otimes R_l^{-1}\ldots \otimes R_l^{-1}\vert \Psi_{\rm sym}\rangle=\langle N/2,-N/2\vert R_l^{-1}\otimes 
R_l^{-1}\ldots \otimes R_l^{-1} \left\{\sum_{k=0}^N\, c_k\, \vert N/2,k-N/2\rangle\right\}=0\nonumber \\ 
 {\rm or} \ \sum_{k=0}^{N}\, c_k\,\langle N/2,-N/2\vert {\cal R}_l^{-1} 
 \vert N/2,k-N/2\rangle= 
 \sum_{k=0}^N\, c_k\  D^{N/2*}_{k-N/2,-N/2}(\alpha_l,\beta_l,0)=0,\hskip 1in  
\end{eqnarray}
\end{widetext}
where we have denoted $R_l^{-1}\otimes 
R_l^{-1}\ldots \otimes R_l^{-1}~=~{\cal R}^{-1}$ in the collective $(N+1)$ dimensional symmetric subspace of $N$ qubits and 
$D^{N/2*}_{k-N/2,-N/2}(\alpha_l,\beta_l,0)=\langle N/2,-N/2\vert{\cal R}^{-1}_l 
 \vert N/2,k-N/2\rangle$,
  represents the collective rotation   
 in the Wigner-$D$ representation~\cite{Rose} 
Substituting the explicit form of the $D$-matrix~\cite{Rose},
%\begin{equation}  
$D^{N/2*}_{k-N/2,-N/2}(\alpha_l,\beta_l,0)=(-1)^k\, \sqrt{^NC_{k}}\, 
\left(\cos\frac{\beta_l}{2}\right)^{(N-k)}\, \left(\sin\frac{\beta_l}{2}\right)^k \, e^{i(k-N/2)\alpha_l}$,
%\end{equation}
in (\ref{cmj0}) and simplifying, we obtain
\begin{eqnarray}
\label{cmj02}
{\cal A}\, \sum_{k=0}^N (-1)^k\,\sqrt{^N\, C_k}\,  c_k\,  \, z^{k}&=&0\,  
\end{eqnarray} 
where $z=\left(\tan\frac{\beta_l}{2}\,e^{i\, \alpha_l}\right),$ and 
${\cal A}=\cos^N\frac{\beta_l}{2}\,e^{-i\alpha_l\, N/2}$. In other words, given the parameters $c_k$, 
the $N$ roots $z_l, l=1,2,\ldots N$ of the Majorana polynomial 
\begin{equation} 
\label{Mp}
P(z)=\sum_{k=0}^N\, (-1)^k\, \sqrt{^N\, C_k}\,  c_k\,  \, z^{k}
\end{equation}
determine the orientations $(\alpha_l,\beta_l)$ of the  spinors 
constituting the  $N$-qubit symmetric state.

\subsection{Classification of pure $N$ qubit symmetric states} 

Two $N$-party pure  states $\vert\phi\rangle$, $\vert \psi\rangle$  can be obtained from one another by means of stochastic local operations and classical communications (SLOCC)~\cite{Dur}  if and only if there exists an invertible local operation 
(ILO) $A_1\otimes A_2\otimes \ldots \otimes A_N$ such that $\vert \phi\rangle=A_1\otimes A_2\otimes 
\ldots \otimes A_N\vert\psi\rangle.$  In the important case of  $N$-qubits obeying permutation symmetry,  it is sufficient 
to search for identical ILOs of the form $A^{\otimes N}=A\otimes A\otimes\ldots \otimes A$ to verify the SLOCC equivalence of 
two pure states~\cite{Solano, Bastin}.   The Majorna representation discussed above leads to a  natural identification of  different entanglement families, depending on the number and arrangement of 
the independent spinors (qubits) constituting the pure symmetric state~\cite{Solano}.  

Let $d\leq N$ be the number of distinct spinors $\vert \epsilon_1\rangle, \vert 
\epsilon_2\rangle,\ldots, \vert \epsilon_d\rangle,$
(which are determined by the independent roots of the Majorana polynomial (\ref{Mp}))
in a $N$ qubit pure symmetric state (\ref{Maj}).  
 The list of degneracy numbers $\left\{n_1,n_2,\ldots n_d;  \sum_i 
n_i=N\right\}$ -- where $d$ denotes the number of distinct spinors  respectively appearing   $n_1\geq 
n_2\geq \ldots n_d$ times, offers a classification of  pure symmetric states into 
different  families~\cite{Solano} denoted by $\{D_{n_1,n_2,\ldots n_d}\}$. 
It may be noted that under identical ILO $A^{\otimes N}$  a symmetric state $\vert D_{n_1,n_2,\ldots, n_d}\rangle$
 belonging to a particular class $\{D_{n_1,n_2,\ldots, n_d} \}$ gets transformed to  
 another state of the same class $\vert D'_{n_1,n_2\ldots, n_d}\rangle=
A^{\otimes N}\vert D_{n_1,n_2\ldots, n_d}\rangle$, 
with the constituent spinors transforming as 
$\vert\epsilon'\rangle~=~A\, \vert\epsilon\rangle$. This forms the main basis of the SLOCC classification of symmetric pure 
states~\cite{Solano}.

For example, when all the $N$ solutions of the Majorana polynomial are identically equal, the symmetric state is given by 
\begin{equation}
\vert D_N\rangle=\vert \epsilon,\epsilon,\ldots \epsilon\rangle,
\end{equation} 
with  $d=1$; the states belong to the family of 
separable  states denoted by $\{D_N\}$.  

The  states with two distinct spinors have the form, 
\begin{eqnarray}
\label{n1n2}
\vert D_{N-k,k}\rangle&=&{\cal N}\,[\vert \underbrace{\epsilon_1, \epsilon_1,\ldots 
\epsilon_1}_{N-k} \underbrace{\epsilon_2, \epsilon_2,\ldots \epsilon_2}_{k}\rangle+{\rm \, Permutations\,}]\nonumber \\
&&  \hskip 0.5in  
\end{eqnarray} 
where $k=0,1,2, \ldots [N/2]$. Dicke states (\ref{Dicke}) are representative states of the entanglement classes 
$\{D_{N-k,k}\}$ and clearly, they 
 are all inequivalent under  SLOCC. 
 
 Further, when the solutions are all distinct, 
 the pure symmetric states constitute  the class $\{D_{1,1,1,\ldots, 1}; \ d=N\}$; the $N$ qubit GHZ state belongs to  this entanglement class.

\section{Determining $N$ qubit states of the family $\{D_{N-k,k}\}$ from its parts}

In this section we show that $N-1$ qubit reduced density matrices uniquely specify the $N$ qubit pure symmetric state belonging to the entanglement class $\{D_{N-k,k}\}$ i.e., no other pure or mixed $N$ qubit state can share the same set of subsystem density matrices.

We first cast the symmetric state (\ref{n1n2}) with two distinct Majorana spinors, belonging to the  family 
$\{ D_{N-k,k}\},$ in  the collective Dicke basis (\ref{Dicke}) as follows: 
%\begin{widetext}
\begin{eqnarray}
\label{dnk}
\vert D_{N-k, k}\rangle &=& {\cal N}\, \sum_{P}\, \hat{P}\,\{ \vert \underbrace{\epsilon_1, \epsilon_1,
\ldots , \epsilon_1}_{N-k};\ \underbrace{\epsilon_2, \epsilon_2,\ldots , \epsilon_2}_{k}\rangle\}\nonumber \\
&=& {\cal N}\, R_1^{\otimes N}\, \sum_{P}\, \hat{P}\,\{ \vert \underbrace{0, 0,
\ldots , 0}_{N-k};\ \underbrace{\epsilon'_2, \epsilon'_2,\ldots , \epsilon'_2}_{k}\rangle\},\nonumber \\
\end{eqnarray}
%\end{widetext}
where we have expressed $\epsilon_1=R_1\vert 0\rangle$ and $\epsilon_2=R_2\vert 0\rangle$, and 
\begin{equation}
\label{ep'}
\vert \epsilon'_2\rangle=R_1^{-1}R_2\vert 0\rangle=d_0\, \vert 0\rangle+d_1\, \vert 1\rangle,\ \ d_0^2+d_1^2=1.
\end{equation}
Substituting (\ref{ep'}) in (\ref{dnk}) and upon simplification, we obtain, 
\begin{eqnarray}
\vert D_{N-k, k}\rangle &=& R_1^{\otimes N}\, \sum_{r=0}^k\,\sqrt{^N C_{r}}\, \alpha_{r}\, \left\vert\frac{N}{2},\frac{N}{2}-r \right\rangle,
\nonumber \\
\label{Ar}
{\rm where} \  \alpha_{r}&=&{\cal N}\,\, 
\frac{(N-r)!}{(N-k)! (k-r)!}\, d_0^{k-r}\, d_1^k. 
\end{eqnarray}
In other words, all symmetric states $\vert D_{N-k, k}\rangle$, constituted by two distinct Majorana spinors are equivalent 
(under local unitary transformations) to 
\begin{eqnarray}
\label{nk'}
\vert D'_{N-k, k}\rangle&=&R_1^{-1\, \otimes N}\vert D_{N-k, k}\rangle \nonumber \\
&=& \sum_{r=0}^k\, \sqrt{^N C_{r}}\, \alpha_{r}\, \left\vert\frac{N}{2},\frac{N}{2}-r \right\rangle.
\end{eqnarray}

We now proceed to prove that {\em parts}  belong {\em uniquely} to the {\em whole} $N$ qubit pure state $\vert D'_{N-k, k}\rangle$.

Let us first express the state $\vert D'_{N-k, k}\rangle$  in the qubit basis: 
\begin{widetext}  
\begin{eqnarray}
\label{dqubit}
|D'_{N-k,k}\rangle&=& \alpha_0\, |0_1,\,0_2,\,\cdots,\,0_N \rangle 
 +\alpha_1 
\sum_{P}\hat{P}\{|1_1,0_2\,\cdots,\,0_{N-1},0_N \rangle\}
   + \alpha_2\, \sum_{P}\hat{P}\{|1_1,1_2\,0_3\,\cdots,\,0_{N}
\rangle\}+\cdots \nonumber \\
 &&+\alpha_k\sum_{P}\hat{P}\{|1_1,\,1_2,\,\cdots,\,1_{k},\,0_{k+1},\cdots,\,0_N
\rangle\}=\vert \phi_0\rangle\, \vert 0_N\rangle+  \vert \phi_1\rangle\, \vert 1_N\rangle
\end{eqnarray}
where
\begin{eqnarray}
\label{phidqubit1}
\vert \phi_0\rangle&=& \alpha_0\ \vert 0_1,\,0_2,\,\cdots,\,0_{N-1} \rangle+\alpha_1\, \sum_{P}\, \hat{P}\,\{\vert 1_1,\,0_2,\cdots,\,0_{N-1} \rangle\}  + \alpha_2\, \sum_{P}\, \hat{P}\,\{|1_1,\,1_2,\,0_3,\,\ldots,\,0_{N-1} \rangle\}, \nonumber \\
&& +\ldots\ldots +\alpha_k\, \sum_{P}\, \hat{P}\,\{|1_1,\,1_2,\,1_3,\,\cdots,\,1_{k},\,0_{k+1},\cdots,\,0_{N-1} \rangle\} \\ 
\vert \phi_1\rangle&=& \alpha_1\ \vert 0_1,\,0_2,\,\cdots,\,0_{N-1} \rangle+\alpha_2\, \sum_{P}\, \hat{P}\,\{\vert 1_1,\,0_2,\,\cdots,\,0_{N-1} \rangle\}  + \alpha_3\, \sum_{P}\, \hat{P}\,\{|1_1,\,1_2,\,0_3,\,\ldots,\,0_{N-1} \rangle\} \nonumber \\
&& +\ldots\ldots +\alpha_{k}\, \sum_{P}\, \hat{P}\,\{|1_1,\,1_2,\,1_3,\,\cdots,\,1_{k-1},\,0_{k},\cdots,\,0_{N-1} \rangle\}.  \end{eqnarray} 

\end{widetext}
Clearly, the $N-1$ qubit reduced density matrix $\rho_{1,2,\ldots, N-1}$ -- obtained by tracing out the $N$th qubit from the state $\vert D'_{N-k, k}\rangle$ --  is a rank-2 mixed state  given by, 
\begin{eqnarray}
\label{nk'2}
\rho_{1,2,\ldots, N-1}&=&{\rm Tr}_{N}[\vert D_{N-k,k}'\rangle\langle 
D_{N-k,k}'\vert]\nonumber \\ 
&=&\vert \phi_0\rangle\langle \phi_0\vert + 
\vert \phi_1\rangle\langle \phi_1\vert, 
\end{eqnarray}

Let us suppose that a mixed  $N$ qubit state $\omega_N$ too shares the same  
$N-1$ qubit reduced system $\rho_{1,2,\ldots, N}$  i.e.,  
\begin{eqnarray} 
 \rho_{1,2,\ldots, N-1}&=&{\rm Tr}_{N}[\vert D_{N-k,k}'\rangle\langle 
D_{N-k,k}'\vert]={\rm Tr}_{N}[\omega_N] \nonumber \\ 
&=&\vert \phi_0\rangle\langle \phi_0\vert + 
\vert \phi_1\rangle\langle \phi_1\vert.  
\end{eqnarray} 
The mixed state $\omega_N$ may always be thought of as a reduced system of a 
{\em pure} state $\vert\Omega_{NE}\rangle$ of the $N$ qubits and an environment $E$ such that 
\begin{equation}
{\rm 
Tr}_{E}[\vert\Omega_{NE}\rangle\langle \Omega_{NE}\vert]=\omega_N.
\end{equation} 
In order that the  pure state $\vert\Omega_{NE}\rangle$ (or the mixed state $\omega_N$) too shares the same $N-1$ qubit reduced density matrix $\rho_{1,2,\ldots, N}$, we must have 
\begin{eqnarray}
\label{wp3}
\vert\Omega_{NE}\rangle&=&\vert \phi_0\rangle\vert E_0\rangle +\vert \phi_1\rangle\vert E_1\rangle, \\ 
\label{on} 
&& \langle E_i\vert E_j\rangle=\delta_{i,j}. 
\end{eqnarray}     
Here, the states $\vert E_{0}\rangle, \vert E_{1}\rangle$ are the ones containing the qubit labelled  $N$,  and the 
environment $E$. Expanding $\vert E_{0,1}\rangle$ in  the basis states of the qubit $N$ as,   
\begin{eqnarray}   
\label{wp4}
\vert E_0\rangle&=&\vert 0_N\rangle\,  \vert e_{00}\rangle+\vert 1_N\rangle\,  \vert e_{01}\rangle \nonumber \\ 
\vert E_1\rangle&=&\vert 0_N\rangle\,  \vert e_{10}\rangle+\vert 1_N\rangle\,  \vert e_{11}\rangle,   
\end{eqnarray}
we may re-express  the state $\vert \Omega_{NE}\rangle$ using (\ref{wp3}), (\ref{wp4}): 
\begin{eqnarray}
\label{omega}
\vert\Omega_{NE}\rangle&=&\vert \phi_0\rangle \vert 0_N\rangle\,\vert e_{00}\rangle +\vert \phi_0\rangle \vert %%@
1_N\rangle\,\vert e_{01}\rangle \nonumber \\
&+& \vert \phi_1\rangle \vert 0_N\rangle\,\vert e_{10}\rangle +\vert \phi_1\rangle \vert 1_N\rangle\,\vert %%@
e_{11}\rangle.
\end{eqnarray} 
If we now demand that yet another $N-1$ qubit reduced system 
$\rho_{2,3,\ldots,N}$ of $\vert D_{N-k,k}'\rangle$  too is shared by 
$\vert \Omega_{NE}\rangle$, it imposes further constraints %~\cite{noteAKR} 
on its structure. It may be readily seen that a comparision of  (\ref{dqubit}) with (\ref{omega}), leads to the identification  
$\vert e_{01}\rangle = 0,$  and  $\vert e_{11}\rangle =\vert e_{00}\rangle$, which 
corresponds to a simplified structure,  $\vert\Omega_{NE}\rangle=\vert D'_{N-k,k}\rangle\, \vert e_{00}\rangle +
\vert \phi_1\rangle\vert 0_N\rangle\vert e_{10}\rangle$.
Now, the orthonormality (\ref{on}) implies that $\langle e_{00}\vert e_{00}\rangle=1$ and $\vert e_{10}\rangle=0$ leading to,  $\vert\Omega_{NE}\rangle=
\vert D'_{N-k,k}\rangle\, \vert e_{00}\rangle$ i.e., the only $N$ qubit state which is consistent with its reduced $N-1$ qubit states is  the state $\vert D'_{N-k,k}\rangle$ itself.  

We verify this result more explicitly by comparing the matrix elements of the $N-1$ qubit reduced density matrix 
 $\rho_{2,\cdots,N}$ of the state $\vert D_{N-k,k}\rangle$  with that obtained from  $\vert \Omega_{NE}\rangle$:  
We first evaluate the following matrix element of the reduced $N-1$ qubit density matrix $\rho_{2,\cdots,N}$
%{\scriptsize$\langle 0_2,0_3\cdots, 0_{N-k-2},1_{N-k-1},\cdots, 1_{N}\vert\rho_{2,\cdots,N}
%\vert 0_2,0_3\cdots, 0_{N-k-2},1_{N-k-1}, 1_N\rangle$},  
\begin{widetext}
\begin{eqnarray}
\langle 0_2,0_3\cdots, 0_{N-k-1},1_{N-k},\cdots, 1_{N}\vert{\rm Tr}_{1,E}[\vert \Omega_{NE}\rangle\langle 
\Omega_{NE}\vert]
\vert 0_2,0_3\cdots, 0_{N-k-1},1_{N-k},\cdots, 1_{N}\rangle&=&\vert\alpha_k\vert^2\, \langle e_{01}\vert e_{01}\rangle, \nonumber \\  
\langle 0_2,0_3\cdots, 0_{N-k-1},1_{N-k},\cdots, 1_{N}\vert{\rm Tr}_{1}[\vert D_{N-k,k}'\rangle\langle 
D_{N-k,k}'\vert]
\vert 0_2,0_3\cdots, 0_{N-k-1},1_{N-k},\cdots, 1_{N}\rangle&=&0.
\end{eqnarray}  
\end{widetext}
from both  
$\vert\Omega_{NE}\rangle$ (see (\ref{omega})) and $\vert D_{N-k,k}'\rangle$  (see (\ref{dqubit})). 
In order to be consistent, we should set  
$\vert e_{01}\rangle\equiv 0$. 
Thus, the state $\vert\Omega_{NE}\rangle$ of (\ref{omega})  
assumes a simpler form, 
%\begin{eqnarray}
%\label{omega2}
\be
\label{Omega2}
|\Omega_{NE}\rangle=|\phi_0\rangle|0_N\rangle |e_{00}\rangle+|\phi_1\rangle\left[|0_N\rangle|e_{10}\rangle+
|1_N\rangle|e_{11}\rangle\right]
\ee%\end{eqnarray} 
Further, from the orthogonality relations (\ref{on}) of the enviornment states 
we get
\be
\label{ortho}
\langle e_{00}|e_{00}\rangle=1,\ \ \langle e_{00}|e_{10}\rangle=0
\ee

Next, we evaluate another matrix element of $\rho_{2,\cdots,N}$ by tracing out the first qubit and the environment 
$E$ of $\vert \Omega_{NE}\rangle$ :   
\begin{widetext}
\begin{equation}
\langle 0_2,0_3\cdots, 0_{N-k},1_{N-k+1},\cdots, 1_{N}\vert{\rm Tr}_{1,E}[\vert \Omega_{NE}\rangle\langle 
\Omega_{NE}\vert]
\vert 1_{2},\cdots, 1_{k+1},0_{k+2},0_3\cdots, 0_{N}\rangle=\vert\alpha_k\vert^2 \langle e_{11}\vert e_{00}\rangle,
\end{equation} 
and compare it with the same matrix element evaluated from $\vert D_{N-k,k}'\rangle$: 
\begin{equation}
\langle 0_2,0_3\cdots, 0_{N-k},1_{N-k+1},\cdots, 1_{N}\vert{\rm Tr}_{1}[\vert D_{N-k,k}'\rangle\langle 
D_{N-k,k}'\vert]
\vert 1_{2},\cdots, 1_{k+1},0_{k+2},0_3\cdots, 0_{N}\rangle=\vert\alpha_k\vert^2.
\end{equation}  
\end{widetext}
This leads to the identification,  $\langle e_{11}|e_{00}\rangle=1,$ which in turn implies (see (\ref{ortho})) that 
$|e_{11}\rangle\equiv|e_{00}\rangle.$ Moreover, the orthonormality relation $\langle E_{10}\vert E_{10}\rangle=
\langle e_{10}\vert e_{10}\rangle+\langle e_{00}\vert e_{00}\rangle=1$ leads to $\vert e_{10}\rangle\equiv 0.$   

In other words, imposing the requirement that the state $|\Omega_{NE}\rangle$ shares  same $N-1$ qubit reduced density matrices  as that of $\vert D_{N-k,k}\rangle$, we  finally obtain,   
\begin{equation}
\label{Omega3A}
|\Omega_{NE}\rangle=|D'_{N-k,\,k}\rangle\, |e_{00}\rangle.
\end{equation} 
In other words, $|D'_{N-k,\,k}\rangle$ happens to be the unique {\em whole} pure state that is consistent with 
its $N-1$ qubit reduced density matrices. We have employed only {\em two} of the $N-1$ reduced density matrices 
$\rho_{1,2,\cdots, N-1}, \rho_{2,3,\cdots, N}$ to establish this result.

\section{Uniqueness of  $N$ qubit states of the generalized family $\{D^{G}_{N-k,k}\}$ with its parts}

The appoach illustrated in Sec.~III suggests a natural extension to a generalized family of $N$ qubit non-symmetric states, obtained from the class $\{ D_{N-k,k}\}$ of states constituted by two distinct Majorana spinors. The generalized class of states have the form, 
\begin{widetext}
\begin{eqnarray}
\label{dgqubit}
|D^{G}_{N-k,k}\rangle&=&\alpha_0\, \, |0_1,0_2,\cdots,0_N \rangle+ \sum_{r=1}^k\, \alpha_r\,\left\{ \sum_{i=1}^{^N C_r}\,a^{(r)}_{i}\, \left[ |1_{P_{(i_1)}},1_{P_{(i_2)}},\ldots, 1_{P_{(i_r)}},0_{P_{(i_{r+1})}}\ldots ,0_{P_{(i_N)}} \rangle\right]\right\} \nonumber \\
&=&\alpha_0\, a_0^{(0)}\, |0_1,0_2,\cdots,0_N \rangle+\alpha_1\,
\left\{a^{(1)}_1\, |1_1,0_2,\cdots,0_N \rangle+a^{(1)}_2\, |0_1,1_2,\cdots,0_N \rangle +\cdots + 
a^{(1)}_N\, |0_1,0_2,\cdots,0_{N-1},1_N \rangle \right\} \nonumber \\
&+& \alpha_2 \,\left\{a^{(2)}_1\, |1_1,1_2,0_3,\cdots,0_N\rangle+a^{(2)}_2\, |1_1,0_2,1_3,0_4,\cdots,0_N \rangle
+\cdots + a^{(2)}_{\frac{(N-1)(N-2)}{2}}\, |0_1,\cdots,0_{N-3},1_{N-2},1_{N-1},0_N \rangle\right.  \nonumber \\ 
&+&  \left. a^{(2)}_{\frac{(N-1)(N-2)}{2}+1}\, |1_1,0_2,0_3,\cdots,0_{N-1},1_N \rangle+\cdots +
a^{(2)}_{\frac{N(N-1)}{2}}\, |0_1,0_2,\cdots,0_{N-2},1_{N-1},1_{N} \rangle\right\}+\cdots \cdots
\nonumber \\
&+&   \alpha_k\,\left\{\, a^{(k)}_1\, |1_1,1_2,\cdots,1_{k},0_{k+1},\cdots,0_N \rangle+\cdots 
+ a^{(k)}_{^{N-1} C_k}\, |0_1,\cdots,0_{N-k-1},1_{N-k},1_{N-k+1},\cdots, 1_{N-1},0_N \rangle\right.\nonumber \\
&+&\left. a^{(k)}_{^{N-1} C_k+1}\, |1_1,1_2\cdots,1_{k-1}, 0_{k},\cdots,0_{N-1},1_N \rangle + \cdots +
a^{(k)}_{^{N}C_k}\, |0_{1},\cdots,0_{N-k},1_{N-k+1},\cdots, 1_N \rangle\right\}.    
\end{eqnarray}
\end{widetext} 
where $\alpha_r$ are as given in (\ref{Ar});  and the states $\sum_{i}\,a^{(r)}_{i}\, \left[ |1_{P_{(i_1)}},1_{P_{(i_2)}},\ldots, 1_{P_{(i_r)}},0_{P_{(i_{r+1})}}\ldots ,0_{P_{(i_N)}} \rangle\right]$  are the generalized Dicke class states with arbitrary coefficients $a^{(r)}_i$. We now proceed to show that there do not exist any other (pure or mixed) $N$ qubit state, which shares the same $N-1$ party subsystem density matrices. 

Expressing (\ref{dgqubit}) in terms of $(N-1,1)$ partition of first $N-1$ qubits and the last qubit i.e.,  
\begin{equation*} 
\vert D^{G}_{N-k,k}\rangle=\vert \phi^{G}_0\rangle \vert 0\rangle_N + 
\vert \phi_1^{G}\rangle \vert 1\rangle_N, 
\end{equation*}
where 
\begin{widetext} 
\begin{eqnarray} 
\label{phig0}
\vert \phi^{G}_0\rangle&=&\alpha_0\, \, |0_1,0_2,\cdots,0_{N-1} \rangle+\sum_{r=1}^k\, \alpha_r\,\left\{ \sum_{i=1}^{^{N-1} C_r}\,a^{(r)}_{i}\, \left[ |1_{P_{(i_1)}},1_{P_{(i_2)}},\ldots, 1_{P_{(i_r)}},0_{P_{(i_{r+1})}}\ldots ,0_{P_{(i_{N-1})}} \rangle\right]\right\} \nonumber \\
&=&\alpha_0\,  |0_1,0_2,\cdots,0_{N-1} \rangle+\alpha_1\,
\left\{a^{(1)}_1\, |1_1,0_2,\cdots,0_{N-1}\rangle+ \cdots + 
a^{(1)}_{N-1}\, |0_1,\cdots,0_{N-2},1_{N-1} \rangle 
 \right\} \nonumber \\
&+& \alpha_2 \,\left\{a^{(2)}_1\, |1_1,1_2,0_3,\cdots,0_{N-1}\rangle+a^{(2)}_2\, |1_1,0_2,1_3,0_4,\cdots,0_{N-1} \rangle
+\cdots + a^{(2)}_{\frac{(N-1)(N-2)}{2}}\, |0_1,\cdots,0_{N-3},1_{N-2},1_{N-1}\rangle\right\}   \nonumber \\ 
&+&\cdots+   \alpha_k\,\left\{\, a^{(k)}_1\, |1_1,1_2,\cdots,1_{k},0_{k+1},\cdots,0_{N-1} \rangle+
a^{(k)}_2\, |1_1,1_2,\cdots,1_{k-1}, 0_{k},1_{k+1},0_{k+2},\cdots,0_{N-1}\rangle \right. \nonumber \\ 
&+& 
 a^{(k)}_3\, |1_1,1_2,\cdots,1_{k-2}, 0_{k-1},1_k,1_{k+1},0_{k+2}\cdots,0_{N-1}\rangle 
+\cdots +a^{(k)}_k\, |0_1,1_2,\cdots,1_{k+1},0_{k+2}\cdots,0_{N-1}\rangle \nonumber \\
&+&\left.  \cdots \cdots +a^{(k)}_{^{N-1} C_k}\, |0_1,\cdots,0_{N-k-1},1_{N-k},\cdots, 1_{N-1}\rangle\right\}, 
\end{eqnarray} 
%\end{widetext} 
and 
%\begin{widetext} 
\begin{eqnarray} 
\label{phig1} 
\vert \phi^{G}_1\rangle&=& \sum_{r=0}^{k-1}\, \alpha_{r+1}\,\left\{ \sum_{i=^{N-1} C_{r+1}+1}^{^{N} C_r}\,a^{(r+1)}_{i}\, \left[ |1_{P_{(i_1)}},1_{P_{(i_2)}},\ldots, 1_{P_{(i_{r})}},0_{P_{(i_{r+1})}}\ldots ,0_{P_{(i_{N-1})}} \rangle\right]\right\} \nonumber \\
&=&\alpha_1 \,a^{(1)}_N\, |0_1,0_2,\cdots,0_{N-1}\rangle+
 \alpha_2 \,\left\{ 
 a^{(2)}_{\frac{(N-1)(N-2)}{2}+1}\, |1_1,0_2,0_3,\cdots,0_{N-1} \rangle+\cdots +
a^{(2)}_{\frac{N(N-1)}{2}}\, |0_1,0_2,\cdots,0_{N-2},1_{N-1} \rangle\right\} 
\nonumber \\
&+& \cdots   + \alpha_k\,\left\{\,  a^{(k)}_{^{N-1} C_k+1}\, |1_1,1_2\cdots,1_{k-1}, 0_{k},\cdots,0_{N-1} \rangle + \cdots +
a^{(k)}_{^{N}C_k}\, |0_{1},\cdots,0_{N-k},1_{N-k+1},\cdots, 1_{N-1} \rangle\right\},  \nonumber \\  
\end{eqnarray}     
\end{widetext}
the $N-1$ qubit reduced density matrix obtained by tracing the $N^{\rm th}$ qubit is readily found to be, 
\be
\label{rhoG}
\rho_{1,2,\cdots, N-1}=|\phi^G_0\rangle\langle\phi^G_0|+|\phi^G_1\rangle\langle\phi^G_1|.
\ee
Suppose that  another $N$ qubit mixed state $\omega^{G}_{N}$ too shares the same $N-1$ qubit reduced system 
(\ref{rhoG}). This requires that an extended pure state $\vert\Omega^{G}_{NE}\rangle$ of $N$ qubits -- appended with an environment $E$ in  such a way that  ${\rm Tr}_E[\vert\Omega^{G}_{NE}\rangle\langle \Omega^{G}_{NE}\vert]=\omega^{G}_{N}$ --  should be expressible as, 
\begin{eqnarray}
\label{OmegaG}
\vert\Omega^G_{NE}\rangle&=&\vert \phi^G_0\rangle\vert E^G_0\rangle +\vert \phi^G_1\rangle
\vert E^G_1\rangle,  \\
\label{onG}
\langle E^G_i\vert E^G_j\rangle&=&\delta_{i,j},
\end{eqnarray}     
in order to be consistent with the marginal state $\rho_{1,2,\cdots, N-1}$.  
Here, $\vert E^G_0\rangle=\vert 0_N\rangle\,  \vert e^G_{00}\rangle+\vert 1_N\rangle\,  \vert e^G_{01}\rangle $ and 
$\vert E^G_1\rangle=\vert 0_N\rangle\,  \vert e^G_{10}\rangle+\vert 1_N\rangle\,  \vert e^G_{11}\rangle $ correspond to the states containing the $N^{\rm th}$ qubit and 
the environment $E$. 
%It is convenient to expand $\vert E^G_0\rangle,\ \vert E^G_1\rangle$ in terms of the basis states of the $N^{\rm th}$ qubit %i.e., 
%\begin{eqnarray}   
%\label{nkEA}
%\vert E^G_0\rangle&=&\vert 0_N\rangle\,  \vert e^{G}_{00}\rangle+\vert 1_N\rangle\,  \vert e^{G}_{01}\rangle \\ 
%\vert E^G_1\rangle&=&\vert 0_N\rangle\,  \vert e^{G}_{10}\rangle+\vert 1_N\rangle\,  \vert e^{G}_{11}\rangle. 
%\end{eqnarray}
Thus the extended  pure state (\ref{OmegaG}) takes  the following form,  
\begin{eqnarray}
\label{Omega2G}
\vert\Omega^G_{NE}\rangle&=&\vert \phi^G_0\rangle \vert 0_N\rangle\,\vert e^G_{00}\rangle +\vert \phi^G_0\rangle \vert
1_N\rangle\,\vert e^G_{01}\rangle \nonumber \\
&+& \vert \phi^G_1\rangle \vert 0_N\rangle\,\vert e^G_{10}\rangle +\vert \phi^G_1\rangle \vert 1_N\rangle\,\vert 
e^G_{11}\rangle. 
\end{eqnarray} 
In order that  $\vert\Omega^G_{NE}\rangle$ shares one more $N-1$ qubit reduced density matrix  
$\rho^{G}_{2,3,\cdots, N}$ (obtained by tracing the first qubit of the original state $\vert D^{G}_{N-k,k}\rangle$) imposes  further constraints which we discuss below.   

We compare the following matrix element of $\rho_{2,3,\cdots, N}$, evaluated from both $\vert\Omega^G_{NE}\rangle$ and $\vert D^{G}_{N-k,k}\rangle$:
\begin{widetext}
\begin{eqnarray}
\langle 0_2,0_3,\cdots,0_{N-1-k},1_{N-k},\cdots 1_N\vert{\rm Tr}_{1,E}\,[\vert\Omega^G_{NE}\rangle
\langle  \Omega^G_{NE}\vert]\vert 
0_2,0_3,\cdots,0_{N-1-k},1_{N-k},\cdots 1_N\rangle &=&|\alpha_k|^2 \vert a^{(k)}_{^{(N-1)}C_k}\vert^2 \langle e^G_{01}|e^G_{01}\rangle,\nonumber \\ 
\langle 0_2,0_3,\cdots,0_{N-1-k},1_{N-k},\cdots 1_N\vert{\rm Tr}_{1}\,[\vert D^G_{N-k,k}\rangle
\langle  D^G_{N-k,k}\rangle\vert]\vert 
0_2,0_3,\cdots,0_{N-1-k},1_{N-k},\cdots 1_N\rangle &=&0  
\end{eqnarray}
\end{widetext}
from which we infer that $\vert e^G_{01}\rangle \equiv 0$. The orthogonality relations (\ref{onG}) would then lead to  
\begin{eqnarray}
\label{onG2}
\langle e^G_{00}\vert e^G_{00}\rangle=1,
%\label{onG3} 
%\langle e_{00}\vert e_{10}\rangle=0.  
\end{eqnarray}

Next, we consider (see Eqs.(\ref{dgqubit})--(\ref{phig1}), (\ref{Omega2G}), 
\begin{widetext}
\begin{eqnarray}
\label{2new}
\langle 0_2,0_3,\cdots,0_{N-k},1_{N-k+1},\cdots 1_N\vert{\rm Tr}_{1,E}\,[\vert\Omega^G_{NE}\rangle
\rangle\langle  \Omega^G\vert]\vert 
1_2,1_3,\cdots,1_{k+1},0_{k+2},\cdots 0_N\rangle
&=&\vert\alpha_k\vert^2   a^{(k)}_{^NC_k} \, a^{(k)*}_k \, \langle e^G_{11}|e^G_{00}\rangle,\nonumber \\
\langle 0_2,0_3,\cdots,0_{N-k},1_{N-k+1},\cdots 1_N\vert{\rm Tr}_{1}\,[\vert D^G_{N-k,k}\rangle
\rangle\langle  D^G_{N-k,k}\vert]\vert 
1_2,1_3,\cdots,1_{k+1},0_{k+2},\cdots 0_N\rangle 
&=&\vert\alpha_k\vert^2   a^{(k)}_{^NC_k} \, a^{(k)*}_k
\end{eqnarray}
\end{widetext}
which imposes the condition 
\begin{equation}
\langle e^G_{11}|e^G_{00}\rangle=1.
\end{equation}
Further, from the orthonormality relation (\ref{onG2}) we obtain,  
\begin{equation}
\vert e^G_{11}\rangle\equiv \vert e^G_{00}\rangle. 
\end{equation}
Consequently, we identify that
%\begin{widetext}   
\begin{eqnarray}
\langle E^G_{10}\vert E^G_{10}\rangle=1&& 
\Rightarrow  \langle e^G_{10}\vert e^G_{10} \rangle+\langle e^G_{00}\vert e^G_{00} \rangle=1 \nonumber \\  
&& \Rightarrow  \langle e^G_{10}\vert e^G_{10}\rangle=0 \ \ {\rm or}\ \ \vert e^G_{10}\rangle\equiv 0. 
\end{eqnarray}
%\end{widetext} 
We are thus  led to our final result 
\be
\label{final}
|\Omega^G_{NE}\rangle=|D^G_{N-k,\,k}\rangle\, |e^G_{00}\rangle
\ee
i.e., the state of the environment is merely a multiple of a fixed basis state 
$\vert e^G_{00}\rangle$ in order that the reduced density matrices of  $|D^G_{N-k,\,k}\rangle$ are shared by 
$|\Omega^G_{NE}\rangle$ too. So, the only {\em whole} $N$ qubit state consistent with its $N-1$ reduced systems is 
the starting state $|D^G_{N-k,\,k}\rangle$ itself.  Only {\em two} of the $N-1$ qubit reduced density matrices 
 suffice to establish this result.   
\section{Summary} 
We have explicitly shown that $N-1$ qubit reduced systems of a pure $N$-qubit symmetric state -- constituted by two distinct spinors --  determine the full state uniquely. No other pure or mixed state of $N$ qubits is consistent with the same set of $N-1$ party marginals. We have employed only {\em two} of the $N-1$ qubit reduced states to establish the uniquenss of the 
{\em whole} pure state  with its {\em parts}. The method developed is readily applied to a generalized family of (non-symmetric)  $N$ qubit states containing two distinct apinors and we have established that specifying only {\em two} of the $N-1$ qubit marginal systems leaves no freedom in the full state to which they belong to. From the point of view of  parametrization of the $N$ qubit state, this suggests that higher order tensor parameters of the whole pure state get completely specified in terms of the lower order tensors. It would be illuminating to explore how the higher order correlations in the whole pure state  originate from the lower order ones. Also, a clear understanding on the  asymptotic convertibility and  reversibility~\cite{Linden} of entanglement within smaller fractions    
in these special classes of $N$ qubit states would shed light on diffierent kinds of multiparty correlations.

\end{document}